\documentclass[aps,prd,showpacs,nobibnotes,twocolumn,preprintnumbers,
nobalancelastpage,superscriptaddress]{revtex4}
\usepackage{latexsym}
\usepackage{dcolumn}
\usepackage{bm}
\usepackage{amsmath}
\usepackage{natbib}
\input{epsf}
\newcommand{\be}{\begin{equation}}
\newcommand{\ee}{\end{equation}}
\newcommand{\ba}{\begin{eqnarray}}
\newcommand{\ea}{\end{eqnarray}}
\newcommand{\bi}{\begin{itemize}}
\newcommand{\ei}{\end{itemize}}
\newcommand{\bfi}{\begin{figure}[!t]
\epsfxsize=9cm
\epsffile}
\newcommand{\bfig}{\begin{figure*}[t]
\epsfxsize=15cm
\epsffile}
\newcommand{\efi}{\end{figure}}
\newcommand{\efig}{\end{figure*}}
\newcommand{\no}{\nonumber}
\newcommand{\la}{\lesssim}
\newcommand{\ga}{\gtrsim}

\begin{document}
\title{Accurate redshift determination of standard sirens by the 
  luminosity distance space-redshift space large scale structure
  cross correlation}
\author{Pengjie Zhang}
\affiliation{Department of Astronomy, School of Physics and Astronomy, Shanghai Jiao Tong
  University, Shanghai, 200240, China}
\email[Email me at: ]{zhangpj@sjtu.edu.cn}
\affiliation{Tsung-Dao Lee institute, Shanghai, 200240, China}
\affiliation{IFSA Collaborative Innovation Center, Shanghai Jiao Tong
University, Shanghai 200240, China}
\affiliation{Shanghai Key Laboratory for Particle Physics and Cosmology, Shanghai 200240, China}
\begin{abstract}
We point out a new possibility to determine the average redshift distribution of
a large sample of gravitational wave standard sirens, without
spectroscopic follow-ups. It is based on the cross correlation
between the luminosity-distance space large scale structure (LSS)
traced by standard sirens, and the redshift space LSS traced by galaxies in preexisting
electromagnetic wave observations. We construct an unbiased and model
independent estimator $E_z$ to realize this possibility. We demonstrate with BBO and Euclid that,
$0.1\%$ accuracy in redshift determination can be achieved. This
method can  significantly alleviate the need of spectroscopic follow-up of
standard sirens, and enhance their cosmological applications. 
\end{abstract}
\maketitle


{\bf Introduction}.---
Gravitational wave (GW) events of black hole (BH)/neutron star
(NS)-BH/NS mergers have been detected
in the nearby universe
\citep{2016PhRvL.116f1102A,2017PhRvL.118v1101A,2017PhRvL.119n1101A,2017PhRvL.119p1101A}
and will be detected in the distant universe by future
experiments. A unique and powerful application of these GW events is
to measure cosmological distance as standard sirens
\citep{1986Natur.323..310S,2017Natur.551...85A}. Such measurement is
based on first principles and therefore  avoid various systematics
associated with traditional methods of electromagnetic (EM) wave
observations.  They will then have profound impact on
cosmology. However, to fully realize this potential, usually it requires
spectroscopic follow-ups to determine redshifts of their host
galaxies or electromagnetic counterparts. This will be challenging for
several reasons. First, some events such as BH-BH mergers may not have
EM counterparts. Second,  future  GW experiments are capable of detecting millions of
standard sirens and the majority of them will be at $z>1$.  EM
follow-ups to determine their spectroscopic redshifts will be 
highly challenging. Various alternatives have been proposed to
circumvent this stringent need of spectroscopic follow-ups
\cite{2016PhRvL.116l1302N,2016PhRvD..93h3511O,2018PhRvD..98b3502N,2018arXiv180806615M}. 

In \cite{Zhang18a} we point out a new possibility. Analogous to the
large scale structure (LSS) traced by galaxies in the redshift space (RS), standard sirens map the
LSS in the luminosity-distance space (LDS). The LSS in this new space by
itself provides the desired redshift information, through the encoded
baryon acoustic oscillation (BAO) and the Alcock-Paczynski test. We
estimate that $1\%$ level accuracy in redshift determination may be
achieved for BBO (the Big Bang Observer, \cite{2006PhRvD..73d2001C,
  2009PhRvD..80j4009C}) or experiments of comparable capability. In the
current paper, we point out that  the LDS-RS LSS cross
correlation can improve the redshift determination accuracy to the
level of $\sim 0.1\%$, yet model independently. We design an estimator
$E_z$, based on  a basic property of LSS. When two LSS (overlapping in sky area)
match better in their redshift distribution, their cross correlations
are tighter. By design, $E_z$ reaches its global
maximum  only when the galaxy redshift distribution 
matches that of standard sirens. Therefore the determined redshift distribution is both model
independent and unbiased. Finding the maximum is essentially a
differential process. Therefore there is a build-in effect of
cancellation of bulk
statistical fluctuations, resulting into S/N higher than 
conventional  estimations. $E_z$ then differs in the above aspects from existing proposals using cross
correlation with galaxies \cite{2016PhRvD..93h3511O,2018PhRvD..98b3502N,2018arXiv180806615M}.
 By the time of the third
generation GW experiments,  there will exist
galaxy surveys of $10^7$-$10^9$ spectroscopic redshifts to $z\sim 1-2$
(e.g. DESI, Euclid, PFS, WFIRST and SKA). 21cm
intensity mapping \cite{2008PhRvL.100i1303C} may probe the even more distant
universe. In combination with them, the average redshift of standard
sirens and its derivative $dz/d\ln D_L$ in many narrow luminosity-distance bins
can be determined to $\sim 0.1\%$ and $\sim 1\%$ accuracy respectively.

{\bf The method}.--- 
Our goal is to determine the true redshift
distribution $\bar{n}^{\rm true}_{\rm GW}(z)$ of standard sirens within a 
luminosity-distance bin ($D_1\leq D_L^{\rm obs}\leq D_2$). $D_L^{\rm obs}$ is the
measured luminosity-distance. It has measurement error of r.m.s.
$\sigma_{\ln D}$, corresponding to r.m.s. redshift error $\sigma_z$.  The average distance is $\bar{D}\equiv
(D_1+D_2)/2$ and the bin width is $\Delta D\equiv
D_2-D_1$.  The true redshifts corresponding to $D_{1,2}$ are
$z_{1,2}$ and the true mean redshift $\bar{z}=(z_1+z_2)/2$. Due to $\sigma_{\ln D}\neq 0$ ($\sigma_z\neq 0$), the true redshift distribution
is wider than $\Delta z\equiv z_2-z_1$. 

For a given galaxy redshift survey, we can apply an arbitrary
weighting function in redshift ($W_{\rm g}(z)$) to form a weighted galaxy
sample. The following $E_z$
estimator measures the mismatch between the galaxy redshifts and
standard siren redshifts, 
\ba
E_z(\ell|W_{\rm g})=\frac{C_{\rm GW-g}(\ell|W_{\rm g})}{\sqrt{C_{\rm g}(\ell|W_{\rm g})}}\ .
\ea
Here $C_{\rm GW}$, $C_{\rm GW-g}$ and $C_{\rm g}$ are the corresponding angular
power spectra respectively. Notice that the cross correlation is measured only
using standard sirens overlapping in sky with the  galaxy survey.  The
expectation value of $E_z$ is $r\sqrt{C_{\rm GW}}$. $r$ is the cross
correlation coefficient between the two LSS. Since $C_{\rm GW}$ is a
fixed quantity, better match in redshift
distribution of the two LSS results into larger $r$ and larger
$E_z$. Therefore the redshift distribution of the  weighted galaxy sample
maximizing $E_z$ tells us  the true redshift  distribution of standard
sirens.  Fig. \ref{fig:Ez} shows the dependence of $E_z$ on the galaxy
redshift distribution. Indeed, when the galaxy distribution has
the same $\bar{z}$ and $\Delta z$ as the standard sirens, its derivatives become
zero and $E_z$ reaches maximum. Notice that $C_{\rm GW-g}$ does not
have this desired property. 

The above argument can be proved more rigorously. The surface
number overdensity of standard sirens and galaxies are 
\ba
\delta_{\rm GW}^{\Sigma}(\hat{\theta})&=&\bar{\Sigma}_{\rm GW}^{-1}\left[\int_{0}^{\infty} \delta_{\rm
  GW}(z,\hat{\theta}) \bar{n}^{\rm true}_{\rm GW}(z) dz\right]\ , \no\\
\delta_{\rm g}^{\Sigma}(\hat{\theta}|W_{\rm
  g})&=&\bar{\Sigma}^{-1}_{\rm
    g}(W_{\rm g})\left[\int_{0}^{\infty} \delta_{\rm g}(z,\hat{\theta})
  \bar{n}_{\rm g}(z) W_{\rm g}(z) dz\right] \ .
\ea
Here $\bar{\Sigma}_{\rm GW}^{-1}=\int_0^\infty \bar{n}^{\rm true}_{\rm
  GW}(z)dz=\int_{D_1}^{D_2} \bar{n}^{\rm obs}(D_L^{\rm obs}) dD_L^{\rm obs}$ is the
mean surface number density of standard sirens.  $\bar{n}_{\rm g}$ is the mean galaxy
number density distribution, fixed by the given spectroscopic redshift
survey. The  weighted galaxy sample has mean surface number density
$\bar{\Sigma}_{\rm g}\equiv \int_{0}^{\infty}  \bar{n}_{\rm  g} W_{\rm g}(z) dz$.   The angular power spectra are 
\ba
C_{\rm GW-g}&=&\frac{\int P_{\rm
    GW-g}(k=\frac{\ell}{\chi(z)},z)\bar{n}^{\rm true}_{\rm GW}\bar{n}_{\rm
    g}W_{\rm g}\chi^{-2}\frac{dz}{d\chi}dz}{\bar{\Sigma}_{\rm GW}\bar{\Sigma}_{\rm
    g}(W_{\rm g})}\ , \no \\
C_{\rm g}&=&\frac{\int P_{\rm g}(k=\frac{\ell}{\chi(z)},z)
  \bar{n}^2_{\rm g}W_{\rm g}^2\chi^{-2}\frac{dz}{d\chi}dz}{\bar{\Sigma}^2_{\rm
    g}(W_{\rm g})}\ .
\ea
$\chi$ is the comoving radial distance. $P_{\rm g}$ and $P_{\rm GW-g}$ are the 3D galaxy and
galaxy-GW host galaxy power spectrum respectively. 
The above expressions adopt a flat universe and the Limber
approximation. But the proof holds otherwise.   Varying $E_z$ with
respect to $W_{\rm g}$, we obtain 
\ba
\delta E_z&=&\int \bar{n}_{\rm
  g}\chi^{-2}\frac{dz}{d\chi}d\chi \times \delta W_{\rm g}(z)\no \\
&&\left(P_{\rm
    GW-g}\bar{n}^{\rm true}_{\rm GW}-P_{\rm g}\frac{C_{\rm GW-g}}{C_{\rm
      g}\bar{\Sigma}}\bar{n}_{\rm g}W_{\rm g}\right)\ .
\ea
The solution to maximize $E_z$ ($\delta E_z/\delta W_{\rm g}=0$) is
\ba
W^{\rm max}_{\rm g}(z|\ell)\propto b_{\rm GW/g}(z)\left(\frac{\bar{n}_{\rm
      GW}(z)}{\bar{n}^{\rm true}_{\rm
    g}(z)}\right) \ .
\ea
Therefore for
    each multipole $\ell$, we have an estimation of the true redshift distribution,
\ba
\hat{n}_{\rm GW}(z)\propto \bar{n}_{\rm g}(z)W_{\rm g}^{\rm max}(z)\propto
b_{\rm GW/g}(z)\bar{n}^{\rm true}_{\rm GW}(z)\ .
\ea
Here $b_{\rm GW/g}(z)\equiv P_{\rm
      GW-g}(k,z)/P_{\rm g}(k,z)$ and $k=\ell/\chi(z)$. In the above
    expressions, we have ignored several normalization factors, since
    the overall normalization is fixed by the 
    total number of observed standard sirens. For the same reason, the
overall amplitude of $b_{\rm GW/g}(z)$ is irrelevant. But its redshift
variation does matter. It biases the estimated  average
redshift by $\delta \bar{z}=b^{'}[(\Delta
z)^2/12+\sigma_z^2]$. Here
$b^{'}\equiv d\ln b_{\rm GW/g}/dz$ at $z=\bar{z}$.  We have verified the excellent agreement between
this  prediction and  the numerical result from the maximum likelihood
fitting  described later.   BBO is able
to achieve $\sigma_{\ln D}\sim 0.02$ ($\sigma_z=0.8\sigma_{\ln D}$ at
$z=1$). This allows us to choose narrow 
luminosity distance bin with $\Delta z\sim 0.04$.  Therefore $\delta \bar{z}\sim 4\times
10^{-4}b^{'}$. If the host
galaxies of standard sirens and EM galaxies are of the same population,
$b^{'}=0$. Otherwise, we expect $|b^{'}|\la 1$ since it may only vary
over cosmic time scale.  Therefore this systematic bias is statistically
insignificant, and will be neglected hereafter.

\bfi{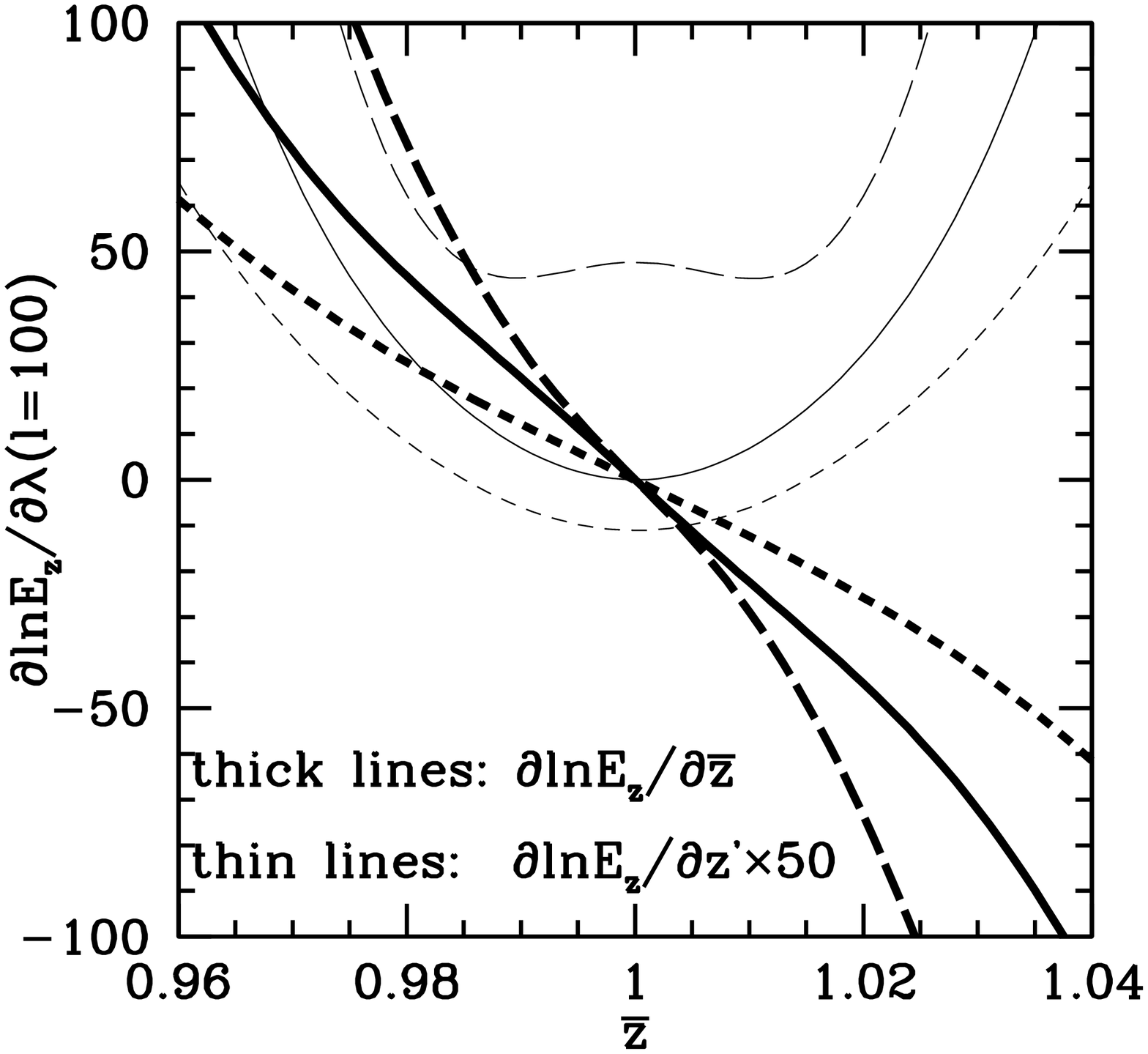}
\caption{The dependence of $\partial E_z/\partial \lambda$ at
  $\ell=100$ on the galaxy redshift distribution,  for fixed standard siren
  distribution with $\bar{z}=1.0$ and  $\Delta z=0.04$. Solid, short
  dash and long dash lines correspond to weighted galaxy samples with 
  $\Delta z=0.04,0.06,0.02$. \label{fig:Ez}}
\efi

Physically, we do not need to vary $W_{\rm g}$ as a completely free
function. The reason is that
there are only limited degrees of freedom in $\bar{n}_{\rm GW}(z)$. It
is fixed by the known PDF $p(D_L|D_L^{\rm obs})$ of distance
measurement error and the $D_L$-$z$ relation to be determined, 
\ba
\label{eqn:nGW}
\bar{n}_{\rm GW}(z)=\int_{D_1}^{D_2}  \frac{dD_L}{dz}p\left(D_L|D^{\rm obs}_L\right) \bar{n}^{\rm obs}(D^{\rm obs}_L) d D^{\rm
  obs}_L \ .
\ea
Since the $D_L$-$z$ relation is smooth, it is naturally described by
the Taylor expansion around $\bar{D}$,
$z(D_L)=\bar{z}+z^{'}(D_L-\bar{D})/\bar{D}+\cdots$. 
Here $z^{'}\equiv dz/d\ln D_L(\bar{D})$.  Given the Taylor expansion
coefficients $\lambda=(\bar{z},z^{'}, \cdots)$, we obtain $\hat{n}_{\rm
  GW}(z|\lambda)$ using Eq. \ref{eqn:nGW}.  Correspondingly,
\ba
W_{\rm g}(z|\lambda)=\frac{\hat{n}_{\rm GW}(z|\lambda)}{\bar{n}_{\rm
    g}(z)}\ ,\ E_z=E_z(W_{\rm g}(z|\lambda))\ .
\ea
Therefore instead of varying a free function $W_{\rm g}$, we only need
to vary a few parameters in $\lambda$. For narrow bins of $\Delta
D/\bar{D}\sim 0.05$ ($\Delta z\sim 0.04$ at $z=1$) that we consider, the Taylor expansion to the linear
order is accurate to $\sim 0.01\%$. Therefore we adopt
$\lambda=(\bar{z},z^{'})$, namely the mean redshift and the slope of
the redshift-distance relation.

{\bf The constraints}.--- 
To avoid model dependence on LSS of standard sirens and galaxies, we do not fit
$E_z$. Instead we only use the model independent condition that $\partial
E_z/\partial \lambda=0$ if the galaxy redshift distribution matches
that of standard sirens. Therefore the (post-processed) data set we will fit is ${\bf 
  D}\equiv \partial E_z/\partial \lambda$.   The corresponding
likelihood is
\ba
p(\lambda|{\bf D})&\propto& p({\bf D}|\lambda)p(\lambda)\propto
\exp\left(-\frac{1}{2}\Delta\chi^2\right)\no \ ,\\
\Delta \chi^2&=&{\bf D}{\bf C}^{-1}{\bf D}^T=\sum_\ell
\left(\frac{\partial E_z}{\partial
    \lambda_\alpha}C^{-1}_{\alpha\beta}\frac{\partial E_z}{\partial
    \lambda_\beta}\right)_{\ell}\ .
\ea
We choose a flat prior on $\lambda$. 
Usually the galaxy number density is orders of magnitude higher
than that of standard sirens. So the covariance matrix
${\bf C}$ is dominated by statistical fluctuations in $C_{\rm
  GW-g}$. It is determined by both statistical fluctuations in the RS
LSS and in the LDS LSS. The former may have comparable contribution from
both shot noise  and cosmic variance in the galaxy
distribution. Therefore we have to keep both. But the later is
dominated by  shot noise,  due to sparse standard siren distribution.
Taking this approximation, we obtain   
\ba
\label{eqn:covariance}
C_{\alpha\beta}&=&\frac{1}{2\ell \Delta \ell f_{\rm
    sky}}\left(\frac{4\pi f_{\rm sky}}{N_{\rm GW}}\right)^2C^{-1}_{\rm
  g} 
\eta_{\alpha\beta}\ , \no \\
\eta_{\alpha\beta}&=&N_{\rm GW}C_{\rm g}\int_0^\infty
\tilde{W}_{,\alpha}\tilde{W}_{,\beta}\bar{n}_{\rm g}dz\no \\
&\times& \left(1+\frac{P_{\rm g}(k=\ell/\chi(z),z)}{4\pi f_{\rm sky}/\bar{n}_{\rm g}}
  \chi^{-2}\frac{dz}{d\chi}\right)\ .
\ea
$\tilde{W}\equiv W_{\rm g}/(\sqrt{C_g}\bar{\Sigma}_{\rm g})$, and $_{,\alpha}\equiv \partial/\partial
\lambda_\alpha$. $\bar{n}_{\rm g}$ has a specific normalization such that
$\bar{n}_{\rm g}(z)\equiv dN_{\rm g}(<z)/dz$ is the number of galaxies per
redshift interval.  $\eta_{\alpha\beta}\lambda_\alpha\lambda_\beta$ is
dimensionless. The second term in the parentheses quantifies the ratio
of cosmic variance and shot noise.

\bfi{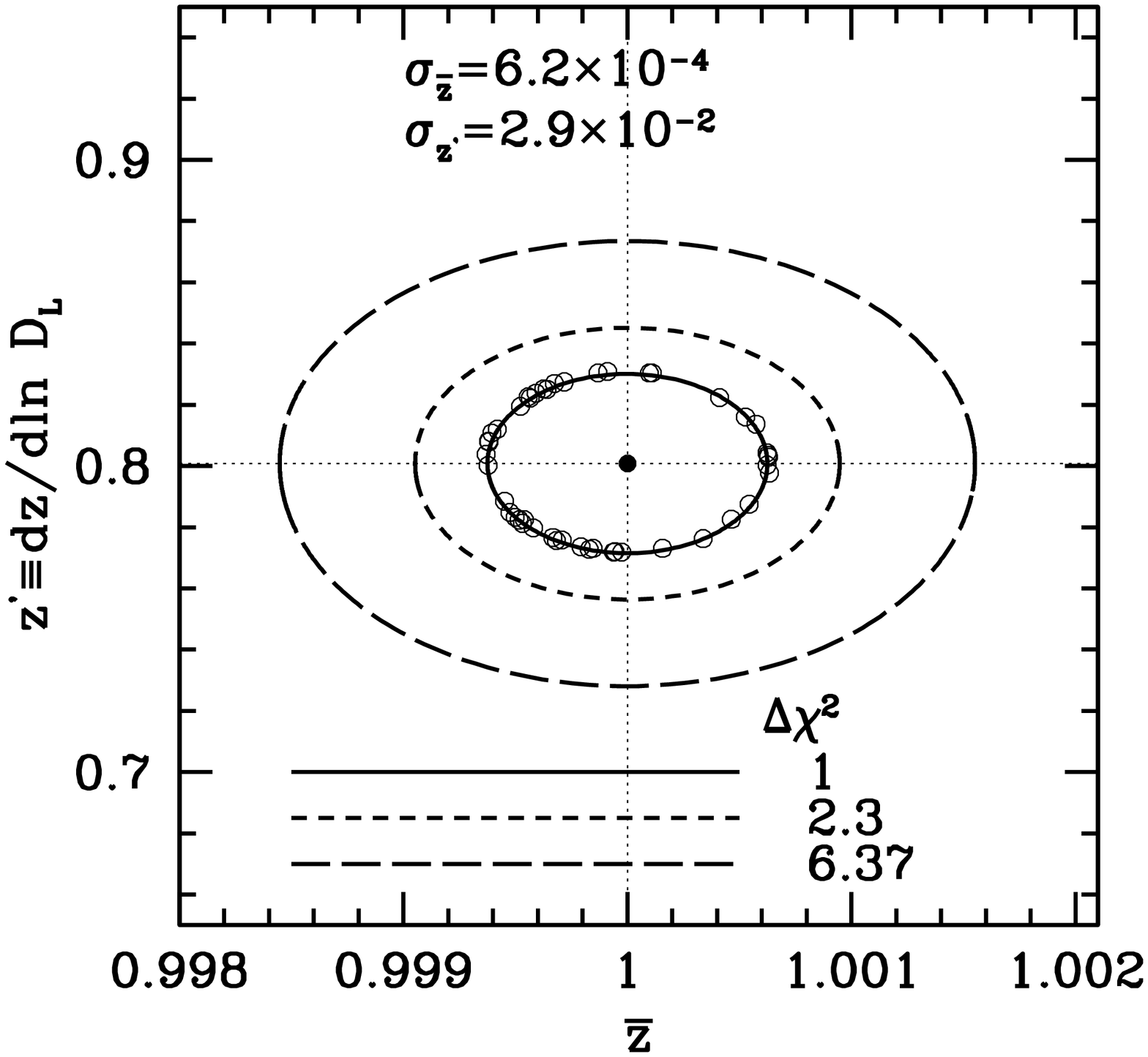}
\caption{The forecast constraints on the mean redshift $\bar{z}$ and
  $z^{'}\equiv dz/d\ln D_L$, with BBO at $\bar{z}=1$ and
  $\Delta z=0.04$ and Euclid. The contours (and $\sigma_{\bar{z},z^{'}}$) are derived from the Fisher
  matrix, and the open circles
  are some random points with the actually calculated $\Delta
  \chi^2\in (0.95,1.05)$. $\Delta \chi^2=2.3 (6.17)$ corresponds
  to $68(95.4)\%$ confidence level. \label{fig:Fisher}}
\efi



We adopt a flat  $\Lambda$CDM cosmology with
$\Omega_m=0.268$, $\Omega_\Lambda=1-\Omega_m$, $\Omega_b=0.044$,
$h=0.71$, $\sigma_8=0.83$ and $n_s=0.96$. We show the forecast on  BBO
\cite{2006PhRvD..73d2001C,2009PhRvD..80j4009C} and Euclid
\cite{Euclid16}, as an example. We follow
\cite{2009PhRvD..80j4009C}  to estimate $\bar{n}_{\rm 
  GW}(z)$, but update the local NS-NS merger rate to a higher value, $R_0=1540 {\rm Gpc}^{-3}{\rm year}^{-1}$
\cite{2017PhRvL.119p1101A}.  The total number of standard sirens per year is  $0.33,1.07,1.77\times 10^6$ at $z<1,2,5$
respectively. The survey duration is adopted as 3 years.  BBO has a
positioning accuracy better than 1 arc-minute \cite{2009PhRvD..80j4009C}. Therefore we will
neglect the angular smoothing of LDS LSS, whose major contribution comes from
$\ell\sim 100$. For $\sigma_{\ln D}$, we adopt $0.02$ \cite{2009PhRvD..80j4009C} as the fiducial
value. But we will also consider the cases of $\sigma_{\ln
  D}=0.01,0.03$. For Euclid, we adopt the galaxy number density as $1.68(0.11)\times 10^{-3} ({\rm Mpc}/h)^{-3}$ at
$z=1(2)$ \cite{Euclid16}. The sky coverage is 15000 deg$^2$ ($f_{\rm
  sky}=0.36$). 

For standard sirens in the bin with $\bar{z}=1$ and $\Delta z=0.04$, 
$\sigma_{\bar{z}}=6\times 10^{-4}$, and $\sigma_{z^{'}}=3\times
10^{-2}$ are achievable (Fig. \ref{fig:Fisher}).  This high accuracy in $\bar{z}$ is
surprising,  given that $C_{\rm GW-g}$ can only
be measured with $\sim 100\sigma$. The reason is that, statistical
fluctuations in $C_{\rm
  GW-g}(\lambda+\delta \lambda)$ and $C_{\rm GW-g}(\lambda)$ are
basically the same in their common redshift
range. Therefore most statistical
fluctuations cancel each in $\partial E_z/\partial
\lambda$.  This cancellation effect is fully captured by $\tilde{W}_{,\alpha}$  in
Eq. \ref{eqn:covariance}, which vanishes near
$\bar{z}$. This is an intrinsic property of the $E_z$ estimator, since
finding the maximum is essentially a differential process. 

Around $z=1$, cosmic variance in the RS LSS mapped by Euclid is comparable to that of shot noise
fluctuations. Therefore including other galaxy surveys helps, but not much.  The errors then scale as
$b_{\rm GW}^{-1}\bar{n}_{\rm GW}^{-1/2}$, where $b_{\rm GW}$
is the density bias of standard siren host galaxies. These errors decrease with decreasing $\sigma_{\ln D}$
(Fig. \ref{fig:zbar}). They also depend on the bin
width $\Delta
D/\bar{D}$, or the equivalent $\Delta z$ (Fig. \ref{fig:zbar}). $\sigma_{z^{'}}$ decreases with increasing $\Delta
z$, for the obvious reason that wider bin size provides better
constraint on the variation of $z$ with respect to $D_L$. In contrast,
$\sigma_{\bar{z}}$ first decreases with increasing $\Delta z$ until
$\Delta z\sim 2.5\sigma_z$, and
then begins to increase with $\Delta z$. This is caused by the
competition of two effects, that larger $\Delta z$ suppresses LSS
information along the radial direction while reducing shot noise.

Combining BBO and Euclid,  the mean  redshift of standard sirens can be determined in many
narrow luminosity distance bins over
$0.7<z<2$.  At $z>1$, the shot noise fluctuation gradually dominates over the cosmic
variance in the galaxy distribution, due to decreasing galaxy number
density. Nevertheless, $\sigma_{\bar{z}}=4\times 10^{-3}$ and
$\sigma_{z^{'}}=0.2$ can still be achieved, for the bin at $\bar{z}=2$ and $\Delta 
z=0.08$. The errors now scale as $\bar{n}^{-1/2}_{\rm g}$. Therefore they
can be significantly reduced by including other surveys such as
PFS \footnote{https://pfs.ipmu.jp/cosmology.html}, the billion 
galaxy survey of SKA2 \cite{2015aska.confE..17A} and WFIRST
\cite{WFIRST13}. The proposed HI intensity mapping by SKA \cite{2015aska.confE..19S} will not only
improve the redshift determination at $z\sim 2$, but also push it to $z\sim 3-4$.

\bfi{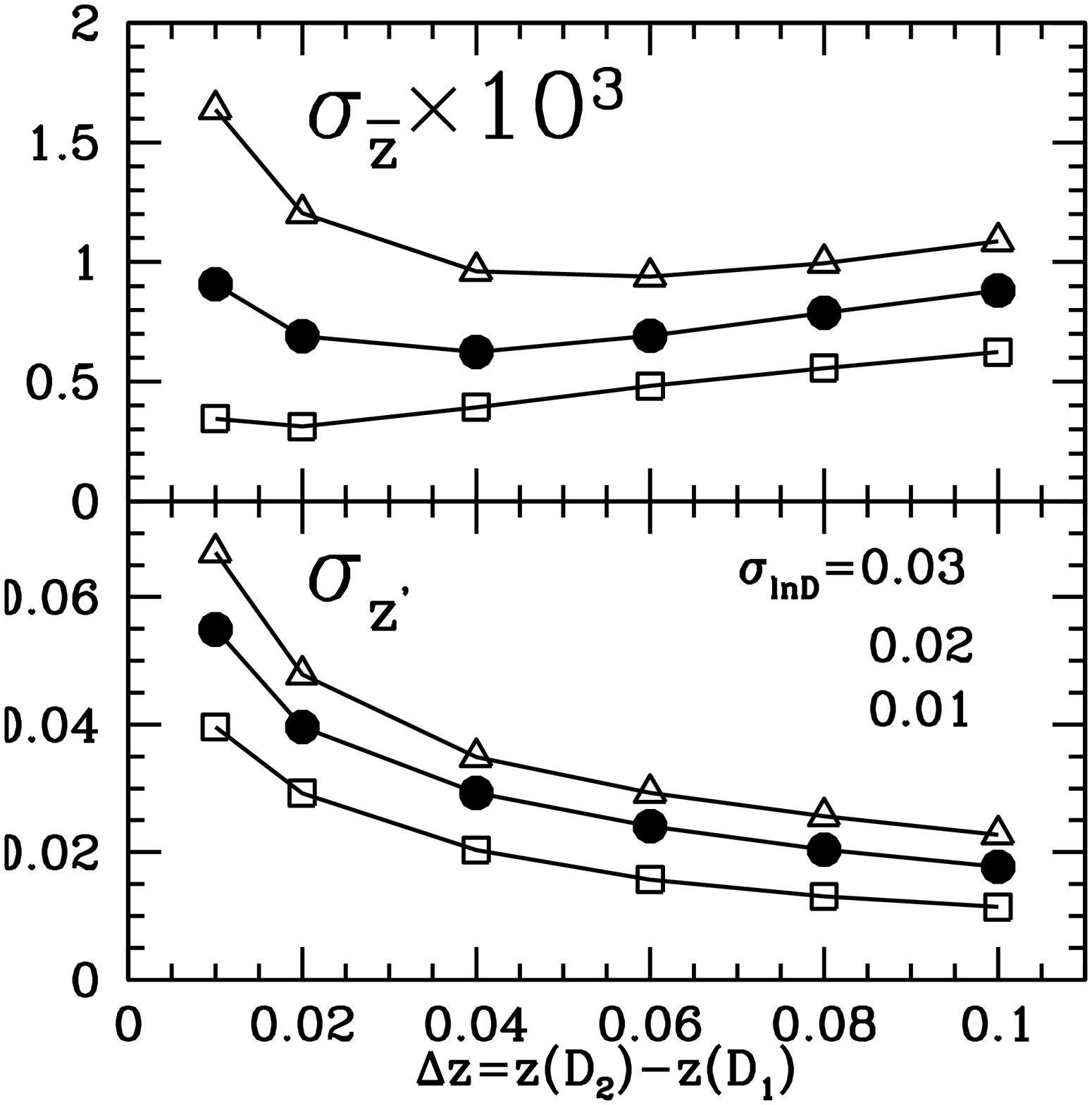}
\caption{The dependence of $\sigma_{\bar{z}}$ and $\sigma_{z^{'}}$ on the bin width
  $\Delta z$. \label{fig:zbar}}
\efi

{\bf Discussions}.--- 
The above proof of concept study neglects several complexities. One is
the lensing magnification on $D_L$. Its direct impact on $\delta_{\rm
  GW}$ is negligible since lensing lacks variation along the
radial separation. However, it increases the effective distance
measurement error ($\sigma^{\rm eff,2}_{\ln D}\simeq \sigma_{\ln
  D}^2+\sigma_\kappa^2$). Since $\sigma_\kappa\sim 0.02$ at $z=1$, it may increase the error
budget by $50\%$ (Fig. \ref{fig:zbar}). Since delensing with cosmic
shear surveys is inefficient \cite{Dalal03}, this may set a lower
limit on $\sigma_{\ln D}^{\rm eff}$, and we may only consider $\Delta z\ga
0.06$. Nevertheless, $0.1\%$ accuracy in $\bar{z}$ is  still
feasible. Another effect neglected is the enhancement of $\delta_{\rm GW}$
and $\delta_{\rm g}$ by coherent peculiar
velocity.  It enhances the effective $b_{\rm GW}$ by $\sim
10\%$ \cite{Zhang18a}, and results into a $\sim 10\%$ reduction in the 
redshift errors. 

The redshift determination achieved by the 
$E_z$ method will allow for many
cosmological applications, beyond the dark energy constraint using the
$D_L$-$z$ measurements. (1) With the accurately determined mean redshift,
the distance duality can be determined to higher 
accuracy than the joint LDS and RS LSS analysis without cross correlation \cite{Zhang18a}. The error will be
dominated by BAO measurement in the RS LSS.  In term of the distance
duality violation parameter 
$\epsilon_a$, BBO and Euclid/SKA are capable of  constraining $\epsilon_a$ to better than $1\%$
over a dozen bins. This will  distinguish between modified gravity models such
as the RR model and GR, with high significance. (2) $z^{'}$ is
analogous to $H(z)$ measured by the
radial BAO of galaxy surveys. It directly tells us the expansion rate
at $z$. It is also a key quantity to break the dark energy-curvature
degeneracy. (3) With the redshift determined, LDS-RS auto and cross
correlations can be combined together to  reduce cosmic variance in constraints of primordial
non-Gaussianity and relativistic effects \cite{2012PhRvD..86f3514Y}.

{\bf Acknowledgement}.---
This work was supported by the National Science Foundation of China
(11621303, 11433001, 11653003, 11320101002), and  National
Basic Research Program of China (2015CB85701).



\bibliography{mybib}
\end{document}